\documentclass[usegraphicx]{mn2e}
\usepackage{subfig}
\usepackage{epsfig}
\usepackage{amsmath}
\usepackage{amssymb}
\usepackage{graphicx}
\usepackage{times,longtable,color}

\usepackage{wrapfig}

\newcommand {\apgt} {\ {\raise-.5ex\hbox{$\buildrel>\over\sim$}}\ }
\newcommand {\aplt} {\ {\raise-.5ex\hbox{$\buildrel<\over\sim$}}\ }

\title[Predicting ULX demographics from geometrical beaming]
{Predicting ULX demographics from geometrical beaming}

\author[M. Middleton \& A. King]
{Matthew J. Middleton$^{1}$ and Andrew King$^{2,3,4}$\\
\\
1. Department of Physics and Astronomy, University of Southampton, Highfield, Southampton SO17 1BJ, UK\\
2. Theoretical Astrophysics Group, University of Leicester, Leicester LE1 7RH\\
3. Anton Pannekoek Institute, University of Amsterdam, Science Park 904, 1098 XH Amsterdam, Netherlands\\ 
4. Leiden Observatory, Leiden University, Niels Bohrweg 2, NL-2333 CA Leiden, Netherlands\\  
}

\pagerange{\pageref{firstpage}--\pageref{lastpage}} \pubyear{2017}
\long\def\symbolfootnote[#1]#2{\begingroup\def\thefootnote{\fnsymbol{footnote}}\footnote[#1]{#2}\endgroup} 
\def\ga{\mathrel{\hbox{\rlap{\hbox{\lower4pt\hbox{$\sim$}}}{\raise2pt\hbox{$>$}}
}}}

\begin{document}

\topmargin = -0.5cm

\maketitle

\label{firstpage}

\begin{abstract} The ultraluminous X-ray source (ULX) population is known to contain neutron stars, but
the relative number of these compared to black hole primaries is unknown. Assuming classical super-critical accretion and resultant geometrical beaming, we show that the observed population ratio can be predicted from the mean masses of each family of compact objects and the relative spatial density of neutron stars to black holes. Conversely - and perhaps more importantly - given even a crude estimate for the spatial densities, an estimate of the fraction of the population containing neutron stars will begin to constrain the mean mass of black holes in ULXs. 
\end{abstract}


\section{introduction}

To first order, super-critical disc accretion is expected to obey classical theory (Shakura \& Sunyaev 1973) with a critical radius at which the mass accretion rate equals the local Eddington limit and the disc inflates towards an aspect ratio of unity due to radiation pressure. The combined loss of material into a wind (see Poutanen et al. 2007; King \& Muldrew 2016) and inwards radial advection (e.g. Abramowicz et al. 1988) can cool the flow, thereby keeping it locally Eddington limited and yields a total radiative luminosity of $\sim L_{Edd}(1+\ln\dot{m}_{0})$ (Poutanen et al. 2007). By itself, this allows even very low mass compact objects such as white dwarfs to appear at luminosities well in excess of $\sim 10^{39}$ erg/s and appear as ultraluminous X-ray sources (ULXs, King 2001) provided the mass transfer rate $\dot{m}_{0}$ (usually quoted in units of Eddington accretion rate, i.e. $\dot{m}_{0} \propto \dot{m}/M$ where $M$ is the compact object mass) is sufficiently high. In the case of high mass binary (HMXB) systems (with $q > 1$) mass transfer eventually shrinks the Roche Lobe of the donor star below the radius of thermal equilibrium such that the donor must expand against the contraction of the Roche Lobe (and thereby return to equilibrium). This expansion drives a period of intense mass transfer on the thermal timescale of the donor (King \& Ritter 1999; King \& Begelman 1999; King, Taam, \& Begelman 2000; Podsiadlowski \& Rappaport 2000) where $\dot{m}\approx\frac{M_*}{t_{KH}}$ (Kolb 1998) ($M_{*}$ is the secondary mass and $t_{KH}$ is the associated Kelvin-Helmoltz time). It is therefore very likely that most HMXB systems will experience periods of super-critical accretion at some stage of their lives.  
 
 \begin{figure*}
\centering
\includegraphics[width=120mm]{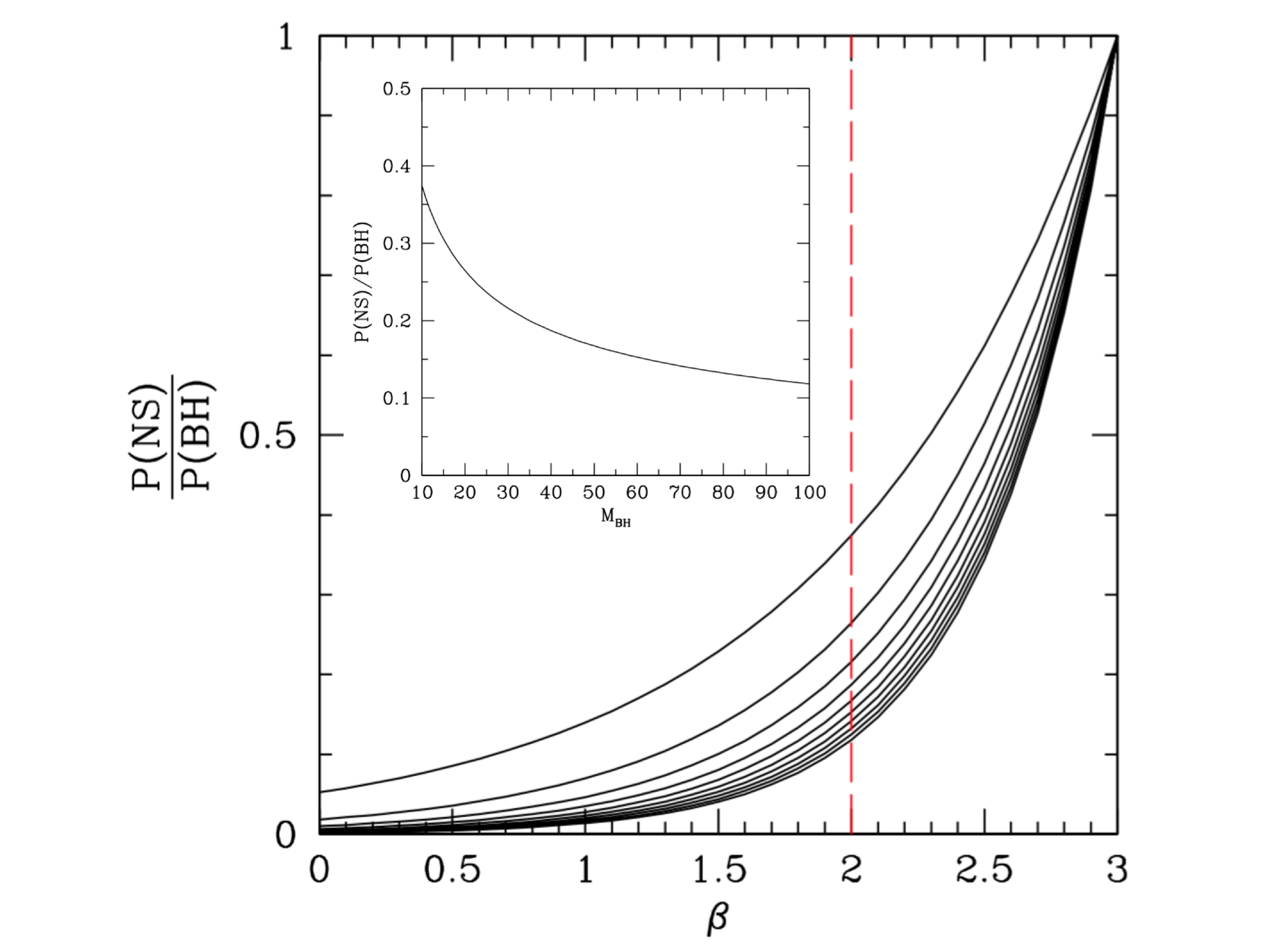}
\caption{Main panel: observed population ratio ($P_{NS}/P_{BH}$) in a flux-limited survey versus beaming index ($\beta$) for a range in black hole mass (10-100$M_{\odot}$ from top to bottom curve) assuming a ratio of neutron star to black hole spatial density of unity, and $M_{NS}$ = 1.4$M_{\odot}$. Whilst the spatial density almost certainly deviates from unity, this is only a multiplicative scaling factor and the overall trend remains unchanged. This demonstrates that the observed population ratio is a relatively steep function of black hole mass (as highlighted in the inset for $\beta$ = 2 - the vertical red dashed line in the main panel); as a consequence, even a rough estimate of the spatial density and observed population will constrain the mean black hole mass in ULXs.}
\end{figure*}
 
The high {\it intrinsic} luminosities in super-critical systems are further amplified as a consequence of the geometry of the flow; the wind launched from the inflated disc is expected to be highly optically thick (Poutanen et al. 2007) leading to an evacuated wind cone from which the majority of the radiation escapes (Ohsuga et al. 2005; Jiang et al. 2014; Sadowski et al. 2014). The result is {\it highly} anisotropic emission with the radiation scattered and beamed towards a favourably inclined observer (King 2009) or de-boosted at higher inclinations (Dauser et al. 2017). Naturally, for smaller wind cone opening angles, the emission becomes increasingly anisotropic and the beaming amplification factor will increase. Geometrical beaming is often cited as an explanation for the most extreme end of the ULX population (above $\approx$1$\times$10$^{40}$ erg/s) and observations charting the evolution in the X-ray spectra and coupled variability appear to match predictions (Middleton et al. 2015). There is additional evidence in favour of geometrical beaming from the apparent lack of eclipses (Middleton \& King 2016) and ULX X-ray luminosity functions in the Local group (e.g. Mainieri et al. 2010), the latter arguing for beaming factors scaling as $\sim \dot{m}_{0}^{2}$ (King 2009). 

In light of the above, the discovery of neutron star primaries in three ultraluminous pulsars (ULPs, Bachetti et al. 2014; Fuerst et al. 2016; Israel et al. 2017a,b) is unsurprising (cf the prediction of King et al. 2001). However, the interpretation of the brightness of these ULPs is still contentious as the nature of the accretion flow will depend heavily on the surface dipole magnetic field strength (see e.g. Mushtukov et al. 2017) and whether this has been effectively diluted by the high $\dot{m}_{0}$ (which itself is not in question). For field strengths $<$ 10$^{12}$G it is probable that most ULPs will be geometrically beamed as the discs reach the Eddington limit before being magnetically truncated (e.g. King \& Lasota 2016). Assuming this is the case, then such beaming must undoubtedly aid in our ability to detect ULXs with either black hole or neutron star primaries out to significantly larger distances than if isotropic emitters. Conversely, as the beaming factor must be tied to the opening angle of the wind-cone, for an isotropic distribution of beam directions, an observer detects a smaller fraction of more tightly beamed sources because of their smaller solid angles on the sky.

In this letter we obtain a simple analytical relationship for the {\it observed} population demographic of neutron stars and black holes in ULXs based on the beaming factor, and which relies on only the mass ratio and the spatial density as free parameters. Note that we do {\it not} assume that all neutron star ULXs show pulsing; as we shall see, most do not.

\section{Analytical estimate from beaming}

We can safely assume that the area of the flux sphere not subtended by the wind ($A$) is inversely related to the level of geometrical beaming, i.e. $b \propto A$ (following the convention laid down in King 2009). The chance probability of detecting a source out to a distance D is therefore given simply by $P \propto nbD^{3}$ where $n$ is the spatial number density of sources with a given primary. Assuming super-critical accretion (and, where the primary is a neutron star, a low surface dipole field strength), a given source luminosity is: 

\begin{equation}
L \propto \frac{L_{Edd}}{b}(1+\ln\dot{m}_0) 
\end{equation} 

We can reasonably assume that the beaming factor is related to the mass transfer rate such that $1/b \propto \dot{m}_0^{\beta} \propto (\dot{m}/M)^{\beta}$ where $\beta$ is some positive valued beaming index. This allows us to write:

\begin{equation}
L \propto  (\dot{m}/M)^{\beta} M(1+\ln\dot{m}_0)
\end{equation}  

\noindent where $L$ is the {\it beamed} luminosity of the source such that in a flux-limited survey (limited to some flux, $f$), $D^{3} \propto (L/f)^{3/2}$. We then observe that: 

\begin{equation}
P \propto  n(M/\dot{m})^{\beta}\left(\frac{\left[(\dot{m}/M)^{\beta} M(1+\ln\dot{m}_0)\right]}{f}\right)^{3/2}
\end{equation}  

The ratio of $P_{NS}/P_{BH}$ is the relative fraction of those ULX primaries found in a flux-limited survey. Assuming that the absolute mass transfer rate is the same for both `species' of ULX, we then find:

\begin{equation}
\frac{P_{NS}}{P_{BH}} = \frac{n(NS)}{n(BH)}\left(\frac{M_{NS}}{M_{BH}}\right)^{(3-\beta)/2}\left(\frac{1+\ln\dot{m}_{0,NS}}{1+\ln\dot{m}_{0,BH}}\right)^{3/2}
\end{equation} 

\noindent where the trailing term is of order unity. This leaves us with:

\begin{equation}
\frac{P_{NS}}{P_{BH}} \approx \frac{n(NS)}{n(BH)}\left(\frac{M_{NS}}{M_{BH}}\right)^{(3-\beta)/2} 
\end{equation} 

Various observational findings (see King 2009) may motivate us to expect $b \propto \dot{m}_{0}^{-2}$ such that:

\begin{equation}
\frac{P_{NS}}{P_{BH}} \approx \frac{n(NS)}{n(BH)}\sqrt{\frac{M_{NS}}{M_{BH}}}
\end{equation} 

\noindent which only depends on the relative spatial densities (also a function of the mass ratio), mean mass of the neutron stars (which covers only a very small range) and the mean mass of the black holes in the ULX sample (assuming Gaussian statistics). 

\section{Discussion \& Conclusion}

We have demonstrated a simple means of predicting the relative observable population of neutron stars and black holes in ULXs from simple beaming arguments that depend on the beaming index ($\beta$), ratio of masses and spatial densities. Although we have reason to believe $\beta \approx$ 2 (King 2009), in Figure 1 we also show the range in observed population ratio ($P_{NS}/P_{BH}$) for a range in beaming index and a range in mean black hole mass (from 10-100M$_{\odot}$) assuming an equal spatial density of neutron stars and black holes, and a canonical neutron star mass (1.4$M_{\odot}$). 

Based on the relative numbers of HMXBs containing neutron stars and black holes (see Casares, Jonker, \& Israelian 2017 and references therein), the true ratio of spatial densities is probably skewed in favour of neutron stars by a factor of $\ge$ 2. Equation 6 then shows that there must be a substantial number of neutron star ULXs even if we do not observe their pulsations (see also the arguments in King, Lasota \& Kluzniak 2017). It is also clear that, unless the true spatial density is {\it heavily} skewed in favour of neutron stars, black hole ULXs still provide a significant (if not dominant) component of the population. This reinforces the argument that the neutron stars in ULXs probably have low to moderate dipole field strengths such that the X-ray spectra do not deviate massively from the remainder of the population (Kluzniak \& Lasota 2015).

The ratio of spatial densities is still somewhat unclear (and will probably require detailed population synthesis). However this only adds a multiplicative scaling factor, so it is immediately clear that for moderate beaming indices, the observed population ratio must be a relatively steep function of black hole mass (see the inset to Figure 1). This opens up the distinct possibility of using flux limited surveys to determine the maximum (mean) mass of black holes in ULXs quite independently of other techniques. 

%add reference to the estimates by 

\section{Acknowledgements}

MJM appreciates support from an Ernest Rutherford STFC fellowship. Astrophysics research at the University of Leicester is supported by an STFC Consolidated Grant.
 
\label{lastpage}

\end{document}